# Enhanced ferroelectricity in epitaxial $Hf_{0.5}Zr_{0.5}O_2$ thin films integrated with Si(001) using $SrTiO_3$ templates

J. Lyu,[1] I. Fina,[1,a)] R. Bachelet,[2] G. Saint-Girons,[2] S. Estandía,[1] J. Gázquez,[1] J. Fontcuberta,[1] and F. Sánchez[1,a)]

[1] Institut de Ciència de Materials de Barcelona (ICMAB-CSIC), Campus UAB, Bellaterra 08193, Barcelona, Spain.

[2] Institut des Nanotechnologies de Lyon (INL-CNRS UMR 5270), Université de Lyon, Ecole Centrale de Lyon, 36 avenue Guy de Collongue, 69134 Ecully Cedex, France.

a) Corresponding authors: ignasifinamartinez@gmail.com, fsanchez@icmab.es

$SrTiO_3$ templates have been used to integrate epitaxial bilayers of ferroelectric $Hf_{0.5}Zr_{0.5}O_2$ and $La_{2/3}Sr_{1/3}MnO_3$ bottom electrode on Si(001). The $Hf_{0.5}Zr_{0.5}O_2$ films show enhanced properties in comparison to equivalent films on $SrTiO_3$(001) single crystalline substrates. The films, thinner than 10 nm, have very high remnant polarization of 34 $\mu C/cm^2$. $Hf_{0.5}Zr_{0.5}O_2$ capacitors at operating voltage of 4 V present long retention time well beyond 10 years and high endurance against fatigue up to $10^9$ cycles. The robust ferroelectric properties displayed by the epitaxial $Hf_{0.5}Zr_{0.5}O_2$ films on Si(001) using $SrTiO_3$ templates paves the way for the monolithic integration on silicon of emerging memory devices based on epitaxial $HfO_2$.

Ferroelectricity in doped $HfO_2$ was recently demonstrated in polycrystalline films arising from a metastable orthorhombic phase (space group $Pca2_1$).[1,2] The phase, not formed in bulk ceramics, is generally stabilized by annealing capped amorphous films. The annealed films are polycrystalline and the orthorhombic (polar) phase coexists with the bulk stable monoclinic (non-polar) phase. The relative amount of phases depends critically on capping, bottom electrode, film thickness, and films thicker than around 20 nm usually present high fraction of monoclinic phase.[2-5] Different factors, including surface energy and strain, have been proposed as responsible of the stabilization of the orthorhombic phase.[5-9] In epitaxial films, where film-substrate interface energy has an important role, the orthorhombic phase has been also stabilized. Remarkably, epitaxial orthorhombic $Hf_{1-x}A_xO_2$ (A = Y or Zr) films have been grown on different oxide substrates: (001)-oriented yttria-stabilized zirconia (YSZ) (bare or indium tin oxide (ITO) coated),[10,12-14] ITO/YSZ(110),[15] ITO/YSZ(111),[11,16] (001)-oriented $La_{2/3}Sr_{1/3}MnO_3$ (LSMO)/$SrTiO_3$ (STO),[17,18, 19] LSMO/$LaAlO_3$(001),[20] and insulating YSZ/Si(001).[21] Integration of epitaxial $Hf_{0.5}Zr_{0.5}O_2$ (HZO) on Si(001) in capacitor geometry has been reported recently using a complex multilayer buffer, LSMO/$LaNiO_3$/$CeO_2$/YSZ/Si(001).[22] A simpler epitaxial structure is desirable, and for this purpose STO can be used as template for the epitaxial integration of ferroelectric $HfO_2$. STO can be deposited by molecular beam epitaxy (MBE), permitting large area deposition and it has excellent compatibility with functional perovskite oxides. Indeed, it is used as buffer layer for deposition of ferroelectric perovskites as $BaTiO_3$ (ref. 23) and $Pb(Zr_xTi_{1-x})O_3$ (ref. 24) on Si(001). The structural compatibility permits epitaxial integration, but the thermal expansion mismatch between the oxides and silicon reduces strongly the out-of-plane polarization of these perovskites respect to equivalent films on $SrTiO_3$(001) single crystalline substrates.[23,24] The effectiveness of STO as buffer layer for epitaxial growth of the orthorhombic phase of $HfO_2$ is unknown. To investigate it, epitaxial STO templates are used here as buffer layers to integrate ferroelectric capacitors with epitaxial HZO films and LSMO bottom electrodes on Si(001). We show that the orthorhombic phase is epitaxially stabilized in films around 8 nm thick and the ferroelectric properties are enhanced respect to equivalent films on perovskite $SrTiO_3$(001) substrates. The remnant polarization ($P_r$) is around 34 $\mu C/cm^2$, its retention longer than 10 years and its endurance against fatigue up to around $10^9$ cycles. Remarkably, both very long retention and high endurance are achieved using same poling voltage.

The epitaxial STO template of thickness t = 26 nm was deposited by solid-source molecular beam epitaxy (MBE) on a Si(001) wafer. Deposition conditions and structural characterization are presented in Supplementary material S1. Bottom LSMO electrodes (t = 25 nm) and HZO films (t = 7.7 nm) were grown on the STO template by pulsed laser deposition (PLD) using a KrF excimer laser. LSMO was deposited at 5 Hz repetition rate, substrate temperature $T_s$ = 675 °C, and dynamic oxygen pressure $PO_2$ = 0.1 mbar. The corresponding growth parameters for HZO were 2 Hz of laser frequency, $T_s$ = 800 °C, and $PO_2$ = 0.1 mbar. The films were cooled under $PO_2$ = 0.2 mbar. The crystal structure was studied by X-ray diffraction (XRD) using Cu Kα radiation. XRD θ-2θ scans, 2θ-χ frames, and ϕ-scans were measured using a Siemens D5000 and a Bruker D8-Advance diffractometers equipped with point and 2D detectors, respectively. Topographic images of the HZO surface were recorded using atomic force microscopy (AFM) in dynamic mode. Platinum top electrodes, 20 nm thick and 20 μm in diameter, were deposited by dc magnetron sputtering through stencil masks for electrical characterization. Ferroelectric polarization loops, retention time, fatigue and current leakage were measured at room temperature in top-bottom configuration using an AixACCT TFAnalyser2000 platform. Polarization loops were obtained at a frequency of 1 kHz using the dynamic leakage current compensation (DLCC) standard procedure.[25,26] Fatigue was evaluated applying bipolar square pulses at a frequency of 100 kHz, whereas staircase triangular pulses of 3 seconds integration time at constant voltage for each data point were used for leakage measurement. Capacitance (C) loops were measured by using an impedance analyzer (HP4192LF, Agilent Co.) operated with an excitation voltage of 50 mV at 20 kHz. Dielectric permittivity ($ε_r$) - voltage loops were extracted from capacitance values using the C= $ε_r ε_0$A/t relation, where A is the electrode area, and t the film thickness.

The XRD θ-2θ scan in Fig. 1a shows (00l) reflections corresponding to the Si wafer, the STO buffer layer and the LSMO electrode. The (111) reflection of orthorhombic HZO (o-HZO), at 2θ around 30°, is accompanied by the small o-HZO(222) peak at around 63°, and no other orientation or phase is observed. The zoomed region in the inset shows the o-HZO(111) peak at 2θ = 30.4° and low intensity Laue interference fringes indicating rather good crystalline and flat interface (see in Supplementary Material, Fig. S2, a XRD scan acquired with longer time and the simulation of the Laue fringes). The out-of-plane interplanar spacing, $d_{o-HZO(111)}$ = 2.940 Å, is smaller to that of equivalent HZO films on LSMO/STO(001) (d = 2.963 Å).[17] The smaller interplanar spacing can be caused by a combined effect of differences in the strain of the LSMO electrodes and in the thermal expansion mismatches between the oxide layers and the Si and STO substrate. Indeed, silicon substrate favors strong in-plane tensile stress to the majority of oxide layers grown at high temperatures.[24] The orthorhombic phase with similar lattice parameter was confirmed in other films on LSMO/STO/Si(001) (Supplementary material, Fig. S3). On the other hand, the XRD 2θ-χ frame (Fig. 1b) shows (001) and (002) reflections of LSMO and STO, and a bright spot corresponding to the o-HZO(111) reflection, whereas the (002) reflection of the monoclinic HZO (m-HZO) phase is barely visible at 2θ around 35°. The narrow intensity distribution of the o-HZO(111) reflection along χ is a signature of highly oriented growth, suggesting that the film is epitaxial. This was confirmed by ϕ-scans around asymmetrical reflections (Fig. 1c). The four (111) reflections of LSMO and STO at 45° of the Si(111) ones signal in-plane rotation of 45° of both layers respect to the Si(001) wafer. The ϕ-scan around o-HZO(-111) shows four sets of three peaks, the 12 peaks being 30° apart. It indicates [1-10]HZO(111) /[1-10]LSMO(001) /[1-10]STO(001) /[100]Si(001) epitaxial relationship and the existence of four HZO crystal domains with in-plane rotation of 90°. The same type of domains were present in HZO films on $SrTiO_3$(001) single crystalline substrates.[17,19] Scanning transmission electron microscopy characterization of an equivalent HZO film and LSMO electrode grown on single crystalline $SrTiO_3$(001) substrate proved the epitaxy and showed the flatness of the HZO/LSMO interface (Supplementary material, Fig. S4). The topographic AFM image (Fig. 1d) shows a flat surface, with root mean square (rms) roughness of 0.35 nm in the 5 µm x 5 µm scanned area. There are some small islands with height up to a few nm, but in most of the topographic image height variations are below 1 nm (see in Fig. 1e the height profile along the horizontal line marked in Fig. 1d).

The current-voltage curve of a Pt/HZO/LSMO/STO/Si(001) capacitor in Fig. 2a shows high amplitude ferroelectric switching current peaks. The remnant polarization in the corresponding polarization loop (Fig. 2b) is around 34 µC/$cm^2$. It is significantly larger than the value of around 20-24 µC/$cm^2$ reached in epitaxial HZO(111) films of similar thickness on LSMO/STO(001).[19] A smaller fraction of monoclinic phase and/or the different lattice strain of the orthorhombic phase could contribute to the larger polarization value. We recently demonstrated that in epitaxial HZO films of similar thickness on STO(001) substrates, the ferroelectric polarization scales with the relative concentration of the orthorhombic phase, whereas the dependence with out-of-plane lattice parameter is less evident.[19] The polarization loop (Fig. 2b) also evidences an imprint voltage towards the negative axis. The imprint voltage of around 0.4 V corresponds to an internal field of around 430 kV/cm pointing towards the bottom LSMO electrode. The large internal field and polarization were confirmed in other HZO films on LSMO/STO/Si(001) (Supplementary material, Fig. S5). On the other hand, the dielectric permittivity loops (Fig. 2c) measured up to a maximum voltage of 4.5 V (loops measured up to larger voltage showed significant leakage contribution), show the butterfly-like shape associated with the ferroelectric nature of the material. Similar loops were observed in the other samples (Supplementary material, Fig. S6), being the permittivity values comparable to those of polycrystalline films.[27,28]

The evolution of the polarization loops when applying high number of bipolar square pulses was investigated. Fig. 3a shows representative loops after different number of cycles of amplitude 5 V. There are not wake-up effects, and the pristine film presents the largest polarization. The memory window reduces with cycling from more than $2P_r$ = 60 µC/cm$^2$ up to around 26 µC/cm$^2$ after 10$^6$ cycles, and further cycling causes capacitor breakdown. When the film is cycled with 4.5 V pulses (Fig. 3b) the initial polarization is smaller, but fatigue is less severe and the capacitor retains memory window larger than $2P_r$ = 16 µC/cm$^2$ after 10$^8$ cycles, before breakdown. Reducing the pulses amplitude to 4 V, the capacitor maintains a significant memory window of $2P_r$ = 4.5 µC/cm$^2$ after 10$^9$ cycles (Fig. 3c). The evolution of the memory window with the number of cycles at 4, 4.5 and 5 V is shown in Fig. 3d. It is remarkable that both endurance and remnant polarization are improved with respect to epitaxial films of similar thickness on perovskite STO(001) substrates.[17-19] The endurance of HfO$_2$-based ferroelectrics is still much smaller that of perovskite ferroelectrics.[29] The endurance up to above 10$^9$ cycles of epitaxial HZO on Si(001) is close to the best reported values for ferroelectric hafnia, obtained in La-doped polycrystalline HZO films.[30]

The endurance study was completed with current leakage measurements after each decade of switching cycles. Representative leakage current versus voltage curves are shown in Supplementary material, Fig. S7. The current leakage at given electric field (300 kV/cm) is plotted against number of switching cycles of different amplitude in Fig. 4. Leakage is below 10$^{-6}$ A/cm$^2$ up to around 10$^3$ cycles for 4, 4.5 or 5V cycling voltage. Above the threshold, leakage increases progressively with additional cycles, and dielectric breakdown occurs when leakage approaches around 10$^{-3}$ A/cm$^2$. It can be seen that for 4V cycling voltage, the leakage remains low until around 10$^7$ bipolar pulses, and dielectric breakdown does not occur until 10$^9$ cycles. Further reduction of leakage current is likely to be possible following doping strategies already applied to polycrystalline ferroelectric HfO$_2$.[30,31]

Ferroelectric retention measurements were performed after poling a capacitor with 5, 4.5 and 4 V (Fig. 5). The reduction of the remnant polarization is small after 10$^4$ seconds, and the extrapolated retention is very high for the three poling voltages. In similar epitaxial HZO films deposited on STO(001) single crystalline substrates retention was also long, but only for large poling fields.[17] In contrast, the retention of Pt/HZO/LSMO capacitor on STO/Si(001) is very high after applying a 4V pulse. To estimate the polarization after longer times, the experimental data are fitted to the power-law decay usually observed[32,33] in ferroelectric oxide films (red dashed curves in Fig. 5). The memory window is reduced by only around 17% of the original value after 10 years (marked with a vertical solid line). Therefore, same poling condition (4V, corresponding to around 4.2 MV/cm) permits good endurance and extraordinarily long retention. Remarkably, these outstanding properties are obtained with pure (undoped) HZO films integrated epitaxially on Si(001) buffered with STO templates, and the properties are improved in comparison to films grown on STO(001) oxide single-crystalline substrates.[17]

In conclusion, SrTiO$_3$ templates permit epitaxial integration of ferroelectric HZO on Si(001). The HZO films, on bottom LSMO electrodes, show high structural quality and enhanced ferroelectric properties with respect to equivalent capacitors on SrTiO$_3$(001) single crystalline substrates. HZO capacitors present large remnant polarization around 34 µC/cm$^2$, and show strong endurance up to 10$^9$ cycles with poling conditions that make the polarization state extremely stable, with extrapolated retention well beyond 10 years. The epitaxial HZO capacitors on SrTiO$_3$/Si(001) platforms can be very valuable to study its intrinsic properties of ferroelectric hafnia, and to develop devices with ultra-small dimensions on semiconductor platforms.

See Supplementary material for Deposition conditions and structural characterization of STO templates (S1), XRD pattern around the o-HZO(111) reflection acquired with longer time and simulation of Laue interference fringes (2), and XRD patterns (S3), cross-sectional transmission electron microscopy image of a HZO/LSMO bilayer on single crystalline STO(001) substrate, ferroelectric polarization loops (S5), and dielectric permittivity loops (S6) of a set of HZO films integrated on Si(001) using STO templates, and leakage current curves measured after cycling the HZO capacitors (S7).

Financial support from the Spanish Ministry of Economy, Competitiveness and Universities, through the "Severo Ochoa" Programme for Centres of Excellence in R&D (SEV-2015-0496) and the MAT2017-85232-R (AEI/FEDER, EU), and MAT2015-73839-JIN projects, and from Generalitat de Catalunya (2017 SGR 1377) is acknowledged. IF acknowledges Ramón y Cajal contract RYC-2017-22531. JL is financially supported by China Scholarship Council (CSC) with No. 201506080019. SE acknowledges the Spanish Ministry of Economy, Competitiveness and Universities for his PhD contract (SEV-2015-0496-16-3) and its cofunding by the ESF. JL and SE work has been done as a part of their Ph.D. program in Materials Science at Universitat Autònoma de Barcelona. INL authors acknowledge the financial supports from the European

Commission through the project TIPS (H2020-ICT-02-2014-1-644453) and from the French national research agency (ANR) through the projects DIAMWAFEL (ANR-15-CE08-0034), LILIT (ANR-16-CE24-0022) and MITO (ANR-17-CE05-0018). They are also grateful to the joint laboratory INL-RIBER and P. Regreny, C. Botella and J. B. Goure for the MBE technical support on the Nanolyon technological platform.

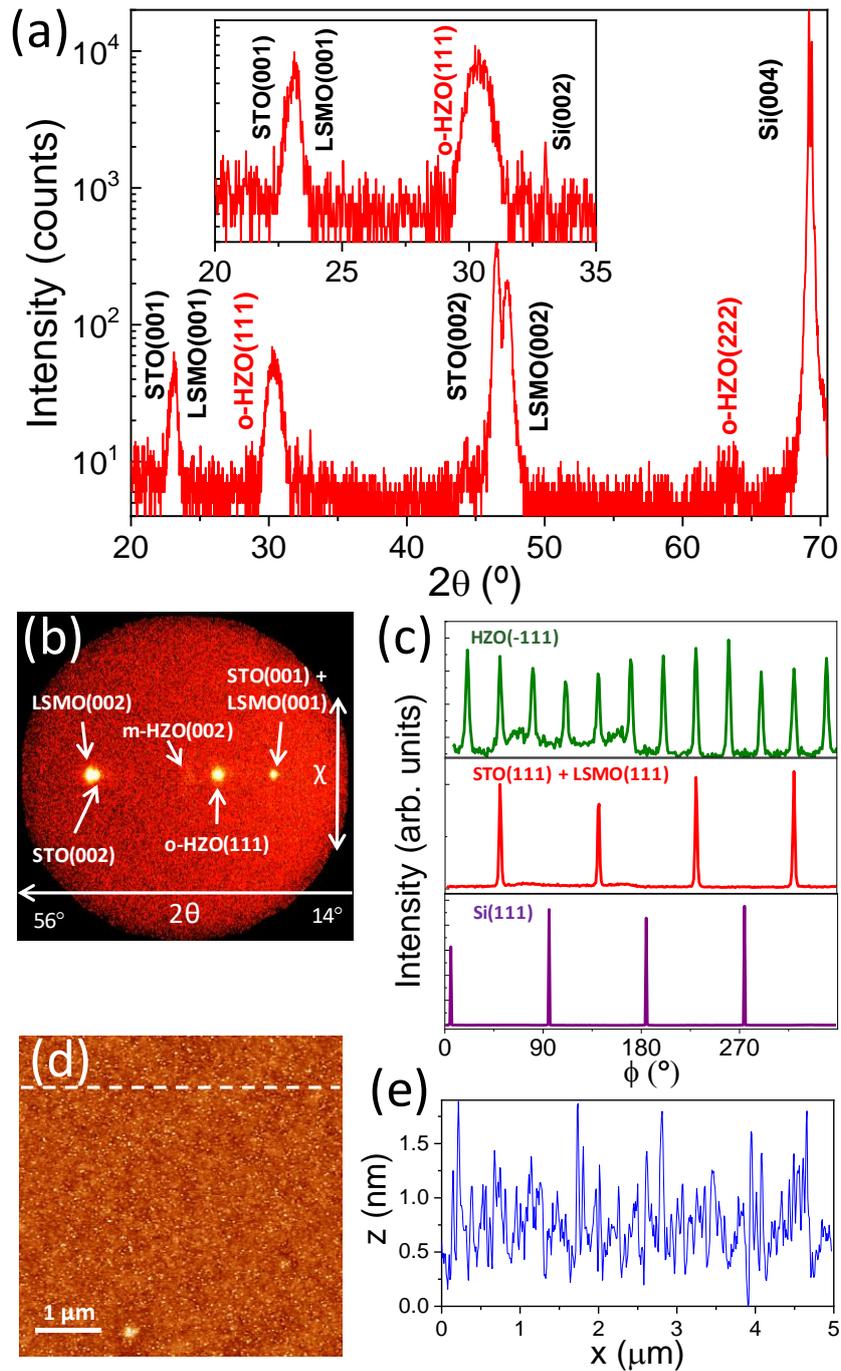

**FIG. 1.** (a) XRD θ-2θ scan of the HZO/LSMO/STO/Si(001) sample. Inset: zoom in the 2θ = 20 - 35° range. (b) XRD 2θ-χ frame around χ = 0°. (c) XRD ϕ-scans around asymmetrical o-HZO(-111), LSMO and STO (111), and Si(111) reflections. (d) Topographic 5 µm x 5 µm AFM image and (e) height profile along the horizontal dashed white line in (d).

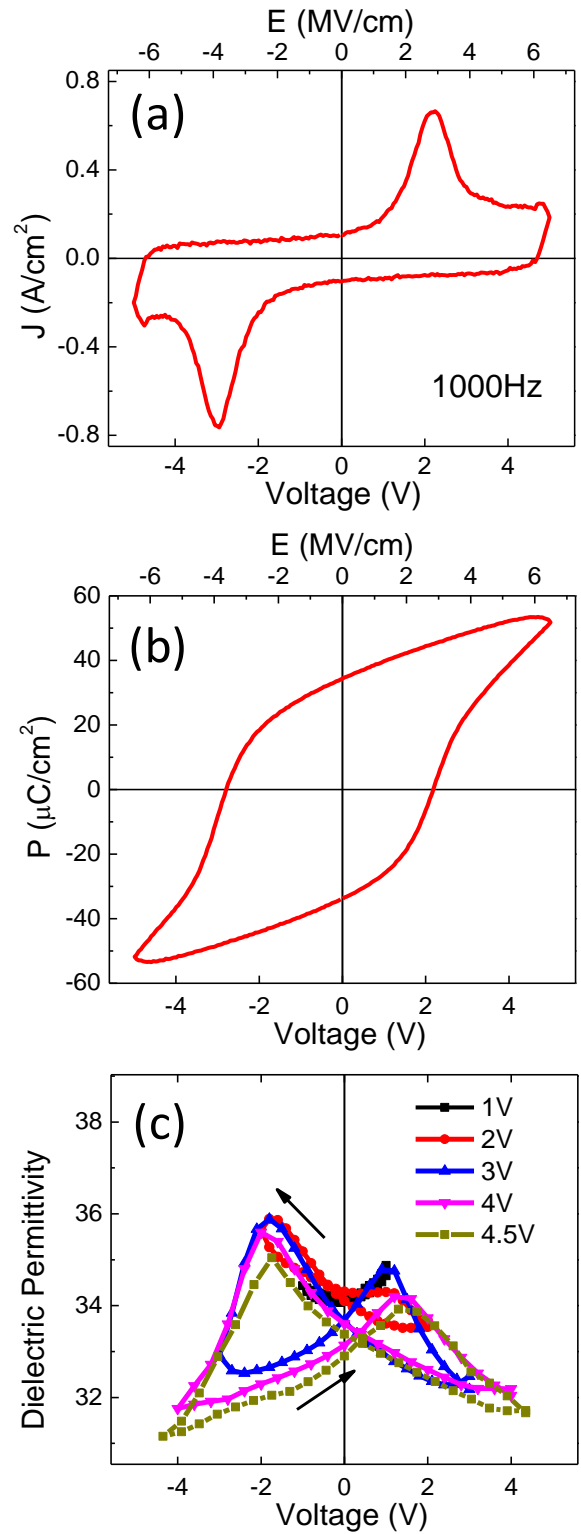

**FIG. 2.** (a) Current-voltage curve and corresponding (b) ferroelectric polarization loop. (c) Dielectric permittivity versus applied voltage loops performed at different maximum applied voltages.

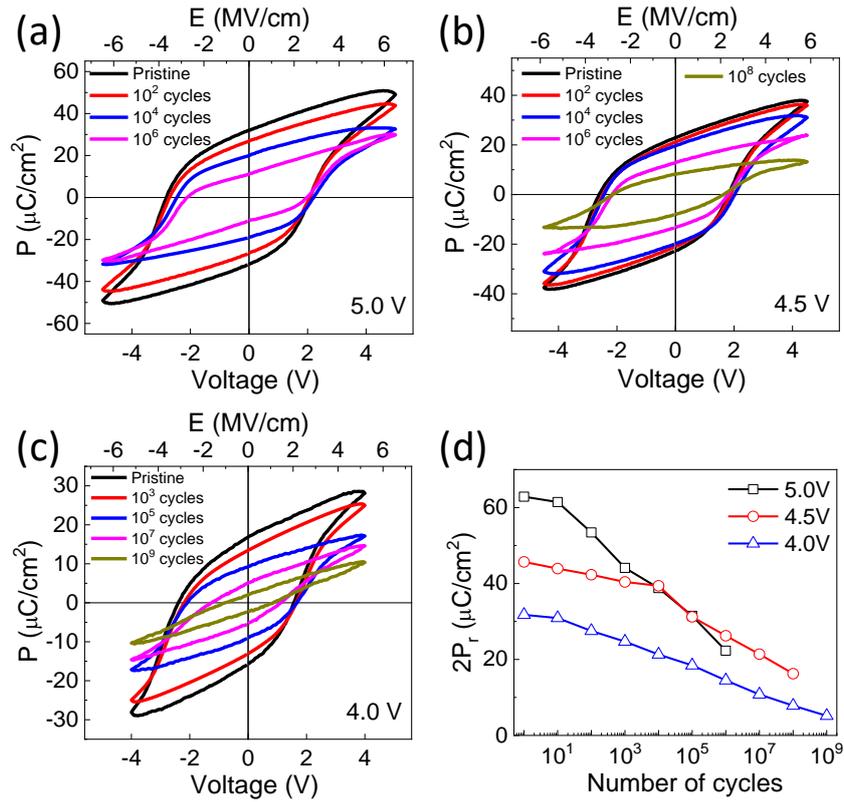

**FIG. 3.** Ferroelectric polarization loops measured at maximum voltage of (a) 5 V, (b) 4.5 V, and (c) 4 V in pristine state and after bipolar switching with the number of cycles indicated in each label. (d) Ferroelectric memory window (2P$_r$) as a function of applied number of cycles at 5, 4.5 and 4 V.

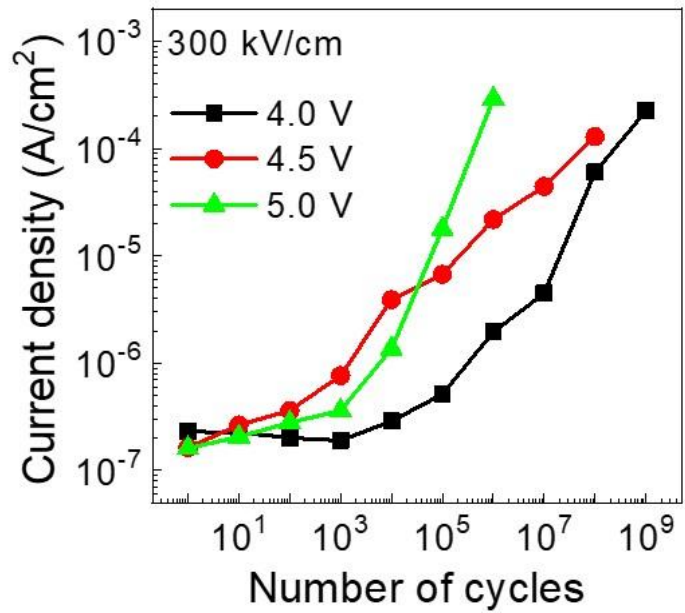

**FIG. 4.** Current leakage at 300 kV/cm as a function of the number of applied cycles of 4, 4.5, and 5 V.

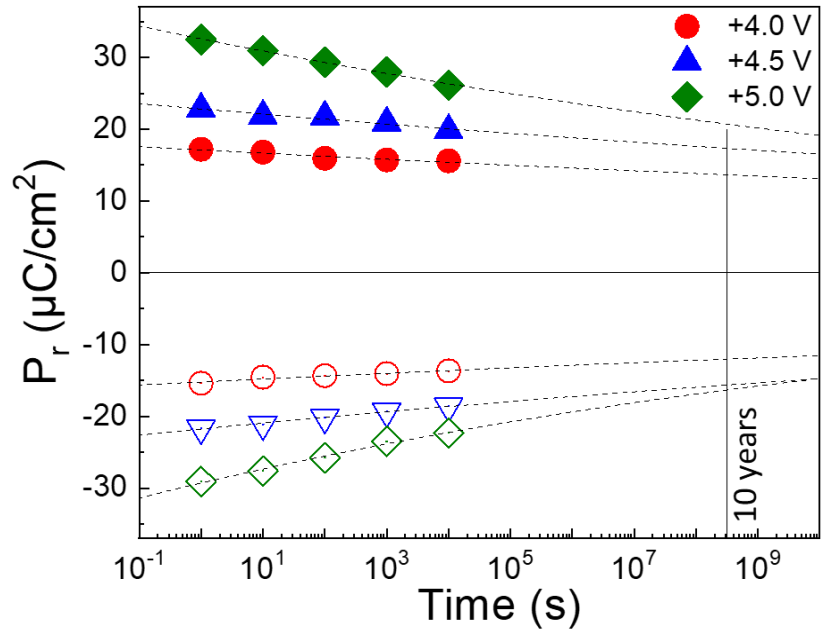

**FIG. 5.** Remnant polarization as a function of the time after poling with a pulse of 5, 4.5 and 4 V. Dashed lines are guides for the eye. Dashed lines correspond to $P_r \propto t^{-n}$ equation data fitting,[32,33] where t refers to the time after poling. The vertical line marks a time of 10 years.

**REFERENCES**

1) T. S. Boscke, J. Müller, D. Bräuhaus, U. Schröder, and U. Böttger, Appl. Phys. Lett. **99**, 102903 (2011).

[2] M. H. Park, Y. H. Lee, H. J. Kim, Y. J. Kim, T. Moon, K. D. Kim, J. Müller, A. Kersch, U. Schroeder, T. Mikolajick, and C. S. Hwang, Adv. Mater. **27**, 1811 (2015).

[3] M. H. Park, H. J. Kim, Y. J. Kim, W. Lee, T. Moon, and C. S. Hwang, Appl. Phys. Lett. **102**, 242905 (2013).

[4] M. Hoffmann, U. Schroeder, T. Schenk, T. Shimizu, H. Funakubo, O. Sakata, D. Pohl, M. Drescher, C. Adelmann, R. Materlik, A. Kersch, and T. Mikolajick, J. Appl. Phys. **118**, 072006 (2015).

[5] M. H. Park, Y. H. Lee, H. J. Kim, Y. J. Kim, T. Shenk, W. Lee, K. Kim, F. P. G. Fengler, T. Mikolajick, U. Schroeder, and C. S. Hwang, Nanoscale **9**, 9973 (2017)

[6] S. E. Reyes-Lillo, K. F. Garrity, and K. M. Rabe, Phys. Rev. B **90**, 140103 (2014).

[7] R. Batra, H. D. Tran, and R. Ramprasad, Appl. Phys. Lett. **108**, 172902 (2016).

[8] R. Materlik, C. Künneth, and A. Kersch. J. Appl. Phys. **117**, 134109 (2015).

[9] E. D. Grimley, T. Schenk, T. Mikolajick, U. Schroeder, and J. M. LeBeau, Adv. Mater. Interf. **5**, 1701258 (2018).

[10] K. Katayama, T. Shimizu, O. Sakata, T. Shiraishi, S. Nakamura, T. Kiguchi, A. Akama, T. J. Konno, H. Uchida, and H. Funakubo, J. Appl. Phys. **119**, 134101 (2016).

[11] K. Katayama, T. Shimizu, O. Sakata, T. Shiraishi, S. Nakamura, T. Kiguchi, A. Akama, T. J. Konno, H. Uchida, and H. Funakubo, Appl. Phys. Lett. **109**, 112901 (2016).

[12] T. Kiguchi, S. Nakamura, A. Akama, T. Shiraishi, and T. J. Konno, J. Ceram. Soc. Jap. **124**, 689 (2016).

[13] T. Shimizu, K. Katayama, and H. Funakubo, Ferroelectrics **512**, 105 (2017).

[14] T. Li, N. Zhang, Z. Sun, C. Xie, M. Ye, S. Mazumdar, L. Shu, Y. Wang, D. Wang, L. Chen, S. Ke, and H. Huang, J. Mater. Chem. C **6**, 9224 (2018).

[15] T. Shimizu, K. Katayama, T. Kiguchi, A. Akama, T. J. Konno, O. Sakata, and H. Funakubo, Sci. Rep. **6**, 32931 (2016).

[16] T. Mimura, K. Katayama, T. Shimizu, H. Uchida, T. Kiguchi, A Akama, T. J. Konno, O. Sakata, and H. Funakubo, Appl. Phys. Lett. **109**, 052903 (2016).

[17] J. Lyu, I. Fina, R. Solanas, J. Fontcuberta, and F. Sánchez, Appl. Phys. Lett. **113**, 082902 (2018).

[18] Y. Wei, P. Nukala, M. Salverda, S. Matzen, H. J. Zhao, J. Momand, A. Everhardt, G. R. Blake, P. Lecoeur, B. J. Kooi, J. Íñiguez, B. Dkhil, and B. Noheda, Nature Mater. **17**, 1095 (2018).

[19] J. Lyu, I. Fina, R. Solanas, J. Fontcuberta, and F. Sánchez, ACS Appl. Electron. Mater. **1**, 220 (2019)

[20] H. Y. Yoong, H. Wu, J. Zhao, H. Wang, R. Guo, J. Xiao, B. Zhang, P. Yang, S. J. Pennycook, N. Deng, X. Yan, and J. Chen, Adv. Funct. Mater. **28**, 1806037 (2018).

[21] K. Lee, T. Y. Lee, S. M. Yang, D. H. Lee, J. Park, and S. C. Chae, Appl. Phys. Lett. **112**, 202901 (2018).

[22] J. Lyu, I. Fina, J. Fontcuberta, and F. Sánchez, ACS Appl. Mater. Interf. **11**, 6224 (2019)

[23] M. Scigaj, C.H. Chao, J. Gázquez, I. Fina, R. Moalla, G. Saint-Girons, M.F. Chisholm, G. Herranz, J. Fontcuberta, R. Bachelet, and F. Sánchez, Appl. Phys. Lett. **109**, 122903 (2016).

[24] R. Moalla, B. Vilquin, G. Saint-Girons, G. Sebald, N. Baboux, and R. Bachelet, CrystEngComm **18**, 1887 (2016).

[25] R. Meyer and R. Waser, Appl. Phys. Lett. **86**, 142907 (2005).

[26] I. Fina, L. Fàbrega, E. Langenberg, X. Martí, F. Sánchez, M. Varela, and J. Fontcuberta, J. Appl. Phys. **109**, 074105 (2011).

[27] M. Park, H. Kim, Y. Kim, T. Moon, K. Kim, Y. Lee, S. Hyun, and C. Hwang, J. Mater. Chem. C **3**, 6291 (2015).

[28] P. D. Lomenzo, Q. Takmeel, S. Moghaddam, and T. Nishida, Thin Solid Films **615**, 139 (2016).

[29] M. H. Park, Y. H. Lee, T. Mikolajick, U. Schroeder, and C. S. Hwang, MRS Commun. **8**, 795 (2018).

[30] A. G. Chernikova, M. G. Kozodaev, D. V. Negrov, E. V. Korostylev, M. H. Park, U. Schroeder, C. S. Hwang, and A. M. Markeev, ACS Appl. Mater. Interf. **10**, 2701 (2018)

[31] U. Schroeder, C. Richter, M. H. Park, T. Schenk, M. Pešić, M. Hoffmann, F. P. G. Fengler, D. Pohl, B. Rellinghaus, C. Zhou, C. C. Chung, J. L. Jones, and T. Mikolajick, Inorgan. Chem. **57**, 2752 (2018).

[32] D. J. Kim, J. Y. Jo, Y. S. Kim, Y. J. Chang, J. S. Lee, J. G. Yoon, T. K. Song, and T. W. Noh, Phys. Rev. Lett. **95**, 237602 (2005).

[33] J. Y. Jo, D. J. Kim, Y. S. Kim, S. B. Choe, T. K. Song, J. G. Yoon, and T. W. Noh, Phys. Rev. Lett. **97**, 247602 (2005).